\documentclass[times]{article}

\usepackage{moreverb}

\usepackage[dvips,colorlinks,bookmarksopen,bookmarksnumbered,citecolor=red,urlcolor=red]{hyperref}

\usepackage[section]{placeins}
\usepackage{flafter}
\usepackage{graphicx}
\usepackage{amsmath}
\usepackage{xspace} 
\usepackage{authblk}

\newcommand\BibTeX{{\rmfamily B\kern-.05em \textsc{i\kern-.025em b}\kern-.08em
T\kern-.1667em\lower.7ex\hbox{E}\kern-.125emX}}

\newcommand{\ijnme}{\textit{International Journal for Numerical Methods in Engineering}\@\xspace}

\newcommand{\prb}{\textit{Physical Review Letters B}\@\xspace}

\renewcommand{\u}{\mathbf{u}}

\begin{document}
\date{}

\title{Inexpensive  discrete atomistic model technique for studying excitations
on infinite disordered media: the case of orientational glass ArN$_2$ \footnote{submitted to \ijnme}}

\author{V.~F.~Gonz\'alez-Albuixech \footnote{Work performed as personal collaboration with the Instituto de Ciencia Molecular.
Current author address: Paul Scherrer Institut. 5232 Villigen PSI, Switzerland
E-mail: vicente.gonzalez@psi.ch} }
\author[1]{A.~Gaita-Ari\~no}

\affil[1]{Instituto de Ciencia Molecular, Universidad de Valencia, Cat. Jos\'e Beltr\'an Mart\'inez n\textordmasculine \ 2 46980, Paterna, Spain}


\maketitle

\begin{abstract}

Excitations of disordered systems such as glasses are of fundamental and
practical interest but computationally very expensive to solve.  Here we
introduce a technique for modeling these excitations in an infinite disordered
medium with a reasonable computational cost.  The technique relies on a 
discrete atomic model to simulate the low-energy behavior of an atomic lattice
with molecular impurities. The interaction between different atoms is
approximated using a spring like interaction based on the Lennard Jones
potential but can be easily adapted to other potentials.  The technique allows
to solve a statistically representative number of samples with a minimum of
computational expense, and uses a Monte-Carlo approach to achieve a state
corresponding to any given temperature. This technique has already been applied
successfully to a problem with interest in condensed matter physics: the solid
solution of N$_2$ in Ar.

\end{abstract}

keywords: glasses; low temperature; atomic modeling; Ar:N2; two-level systems; disordered lattices, universality



\vspace{-6pt}

\section{Introduction}
\vspace{-2pt}

Excitations of disordered systems such as glasses are computationally very
expensive to solve. The first difficulty arises because molecular, atomic and
quantum effects are so important that one needs an atomistic description to
understand the origin of the thermal and mechanical properties at low
temperature.  Nevertheless, and contrary to the situation in ordered crystals,
it is not merely difficult to calculate the quantum ground state but there is
actually no real physical ground state that the system can reach experimentally
at low enough temperatures. Instead, there is just a complicated Potential
Energy Landscape where the low-energy state depends on the thermal history if
the system. It is also impossible to apply real periodic boundary
conditions, as the disorder itself is not periodic, so border effects have to
be suppressed by using large fragments. What is worse, a critical parameter
that determines the properties of such systems is its composition, so the study
needs to be performed at each concentration of impurities. 

Disordered systems actually quite common, so the problem is of general interest.
Indeed, for many real solids progressively lowering the temperature produces an
apparently infinite series of ever more weak interactions, and solid-state
physicists do not presently understand the basic structure of the low-energy
states of most systems, excluding the most simple ones.

In this context, a Two Level System (TLS) corresponds formally to a related
pair of local minima, or a double-well potential on the Potential Energy
landscape (PEL). These minima need to have an small energy difference and
distance to allow for tunneling ~\cite{reinisch}. TLSs are believed to be the
origin of certain {\it universal} properties in disordered solids, and also the
cause of $1/f$ noise in superconducting qubits, a bottleneck in Quantum
Technologies. Thus there is interest in studying them; a crucial step being, of
course, determining their nature, ~\cite{moshe,alex, poh,churkin}. For the reasons
aforementioned, computer simulations are strongly limited and usually only
rather small systems are used to analyze the PEL and TLS properties. This may
give rise to significant finite size effects and may strongly influence the
properties of the TLS obtained by computer simulations.  The computational cost
also imposes limitations on the possibility of obtaining enough results to have
statistic significance, affecting the validity of the results. 

In this work we introduce a technique for modeling TLS that allows to model a
infinite disordered medium  with a reasonable computational cost.  The
technique relies on a discrete atomic model to simulate the low-energy behavior
of an atomic lattice with impurities (displaying TLSs).
The discrete model is based on the equilibrium equations of atomic interaction
forces ~\cite{kwon,burc} as the equilibrium state of the lattice corresponds to
the minimal value of the total potential energy of the atomic structure and
uses a Monte-Carlo method to set the working temperature. As the numerical formulation is 
similar to other well known classical formulations, like the finite element method, 
the introduction of boundary conditions is simplified.  The density
of impurities can thus be iteratively increased up to the desired level by
adding impurities at random positions. A number of independent histories are
stored to achieve a statistically representative set of random configurations.

To illustrate the method, we focus our study on the solid solution of the
nitrogen molecule N$_2$ in an Ar lattice, where the TLSs are expected to be
closely related to different orientations of the N$_2$ molecules. As no
relevant three-dimensional-specific effects are expected in this system, for
ease of visualization we describe here only the study in two dimensions.
Displaying only weak van der Waals forces and being an intriguing system in the
context of universality of low-temperature properties of disordered systems,
Ar:N$_2$ is particularly well-suited for exemplifying our approach, which is
extensible to many systems of different nature. The present work deals mainly
with the new methodology, while the physical consequences of the results we
obtain are presented elsewhere~~\cite{gaitavi}.

\vspace{-6pt}

\section{Potential description for the discrete atomic model}
\vspace{-2pt}

We use a discrete atomic model to compute and simulate the equilibrium of an
atomic Ar lattice including N$_2$ impurities.  The equilibrium state of the
lattice corresponds to the minimal value of the total potential energy of the
atomic structure. This model is based on the equilibrium equations of atomic
interaction forces ~\cite{kwon,burc}.  The model assumes interactions between
each atomic pair which are approximated using a non linear spring
model~\cite{burc,wang}, even if the potential energy can be described using
different equations depending on distance between each 2, 3 or many atoms.  For
sake of simplicity, we based the approach on the Lennard Jones potential
~\cite{lejo}, as it is one of the most extensively used for fluids and solids
~\cite{karakasidis,verlet1,verlet2} and also for large systems ~\cite{heyes}.

The Lennard Jones 6-12 function is used to describe the interaction between two 
atoms of the lattice. The potential energy of the Lennard-Jones function is expressed as
\begin{equation}
\label{eq:V-lj-1}
\begin{aligned}
V_{ij}(r_{ij})=4 \varepsilon\left[ \left(\frac{\sigma}{r_{ij}}\right)^{12}-\left(\frac{\sigma}{r_{ij}}\right)^{6}\right]
\end{aligned}
\end{equation}
\noindent where $\varepsilon$ denotes the well depth and $\sigma$ the
zero-potential distance between two atoms. $r_{ij}$ is the existing distance
between the interacting atoms $i$ and $j$. It is well known that the
Lennard-Jones interactions decrease rapidly as distance increases, the
potential becoming negligible if the distance $r_{ij}$ is much greater than
zero-potential distance. The usual choice for the cut-off distance, which we
adopt, is 2.5$\sigma$.  

The former description corresponds to the existing interaction between two Ar,
or between a Ar and N atom. The potential parameters (energies in Hartree,
distances in Bohr radius) are $\varepsilon_{Ar-Ar}=3.7936*10^{-4} eH$,
$\sigma_{Ar-Ar}=6.3302 r_B$, ~\cite{nielaba,hoover} ,
$\varepsilon_{Ar-N}=2.1263*10^{-4} eH$, $\sigma_{Ar-N}= 6.3306 r_B$
~\cite{nielaba,raman}.

\begin{figure} [htb]
\begin{center}
		\includegraphics[width=0.70\textwidth]{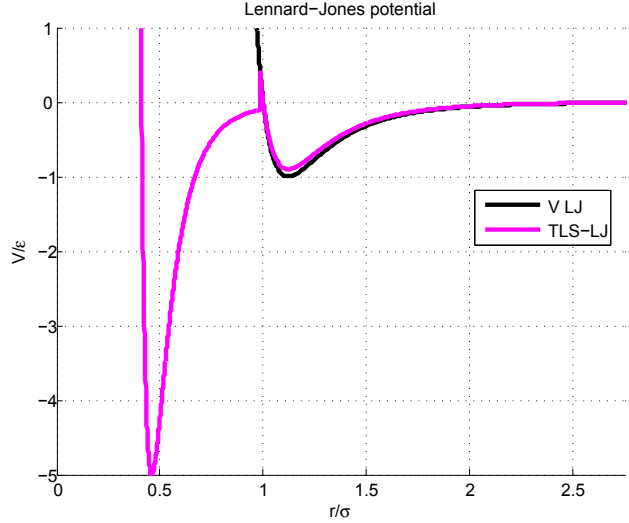}
	\caption{Qualitative illustration of the potentials between Ar atoms (V LJ) and double potential to reproduce both the covalent bond and the weaker interactions of N$_2$ molecules using the Lennard-Jones potential (TLS-LJ). For clarity we choose $\sigma^{TLS}=\sigma$, $\varepsilon^{TLS}=0.9\varepsilon$, $\sigma^{TLS}_{W}=0.41\sigma$, $\varepsilon^{TLS}_W=10\varepsilon$. Of course, in the actual calculations the potential for the covalent bond would be much deeper.}
	\label{fig:pote-tls}
\end{center}
\end{figure}

The N-N interaction deserve some extra cautions, as we need to reproduce both
the intramolecular behavior of the N$_2$ molecule, with a very stiff and rather
short covalent bond, and the intermolecular interaction between two molecules,
which is weak and similar to the Ar-Ar interaction. We do this by using a
potential composed by two very different LJ functions. Of course, the
transition between the two potentials has to be chosen carefully to avoid
numerical instabilities or mathematical artifacts. With these considerations
the potential for N-N interactions is chosen as:
\eqref{eq:V-TLS}
\begin{equation}
\label{eq:V-TLS}
\begin{aligned}
V^{TLS}_{ij}(r_{ij})=4  \left\{ \begin{array}{ccc}\varepsilon^{TLS}\left[ \left(\frac{\sigma^{TLS}}{r_{ij}}\right)^{12}-\left(\frac{\sigma^{TLS}}{r_{ij}}\right)^{6}\right] & & \text{region 1}\\
\\
\varepsilon^{TLS}_{W}\left[ \left(\frac{\sigma^{TLS}_{W}}{r_{ij}}\right)^{12}-\left(\frac{\sigma^{TLS}_{W}}{r_{ij}}\right)^{6}\right] & & \text{region 2}  \end{array} \right.
\end{aligned}
\end{equation}
\noindent region 1 indicates the region outside the bond link of the N$_2$ molecule and regions 2 indicates the bond affected region inside the N$_2$ molecule, which produces the potential well and also the TLS effect, superscript $TLS$ remarks the TLS association, subscript $W$ indicates the parameters for this inner potential well which is considered deeper. The region 2 definition is controlled using a cut off function in function of $\sigma^{TLS}_{W}$, regions 1 starts after this cut off for regions 2; beyond the cut off distance $2.5\sigma_{TLS}$ the potential is defined as zero. Considering the Ar:N$_2$ mixture, we will have the following parameter   $\varepsilon_{N-N}=\varepsilon^{TLS}_{W}= 0.3601 eH$, $\sigma_{N-N}=\sigma^{TLS}_{W}= 2.0749 r_B$, ~\cite{raman},
$\varepsilon_{N_2-N_2}=\varepsilon^{TLS}=1.3916*10^{-4} eH$, $\sigma_{N_2-N_2}=\varepsilon^{TLS}= 6.3136 r_B$
,~\cite{johnson}; the inner cut-off distance is $2.43\sigma_{N-N}$ that is where region 2 ends. The qualitative behavior of potentials for the typical interatomic/intermolecular vs intramolecular potential is shown in Fig. \ref{fig:pote-tls}.

\vspace{-6pt}

\section{Spring-like description for the discrete atomic model}
\vspace{-2pt}

Our model assumes spring like bonded atoms on static equilibrium. That is, the
interaction between each atomic pair can be approximated using a spring model as
in ~\cite{burc,wang}. More exactly, the Lennard-Jones bond behavior is
approximated as a non-linear spring: 

\begin{equation}
\label{eq:spr-lj}
\begin{aligned}
\textbf{f}_{ij} =\textbf{k}_{ij}(\textbf{u}_i-\textbf{u}_j) 
\end{aligned}
\end{equation} 

The interaction forces between each pair of atoms in the lattice with non zero
potential are computed from the potential function, that is: 
\begin{equation}
\label{eq:f-lj-1}
\begin{aligned}
\textbf{f}_{ij} (r_{ij})=-\frac{\partial V(r_{ij})}{\partial r_{ij}}=\frac{4\varepsilon}{r_{ij}} \left[ \left(-12\frac{\sigma}{r_{ij}}\right)^{12}+6\left(\frac{\sigma}{r_{ij}}\right)^{6}\right]\textbf{n}_{ij}
\end{aligned}
\end{equation} 

However in practice we will use the distance vector between the constitutive
atoms of the pair $\textbf{r}_{ij}$ instead of the direction, as in ~\cite{burc}
to achieve the following expression:
\begin{equation}
\label{eq:f-lj-2}
\begin{aligned}
& \textbf{f}_{ij}(r_{ij})=f_{ij}\frac{\textbf{r}_{ij}}{r_{ij}} \\
& \textbf{f}_{ji}=-\textbf{f}_{ij} \\
& f_{ij}=\frac{4\varepsilon}{r_{ij}} \left[ \left(-12\frac{\sigma}{r_{ij}}\right)^{12}+6\left(\frac{\sigma}{r_{ij}}\right)^{6}\right]
\end{aligned}
\end{equation}

We inferred $\textbf{k}_{ij}$ from the slope of the force. However, if
distance $r_{ij}$ is greater than 1.244 $\sigma$, the slope is negative. The
introduction of a negative stiffness may induce some numerical problems and,
for our case, lacks of physical interpretation, hence the absolute value of the
slope is used. No special treatment is done for the special case of
$1.244\sigma$ where the slope is zero, as the possibility of it becoming a
explosive point is almost negligible. The Exact expression for the stiffness
is then
\begin{equation}
\label{eq:k-lj}
\begin{aligned}
&\textbf{k}_{ij} (r_{ij})=\left|k_{ij}\frac{\textbf{r}_{ij}}{r_{ij}}\right| \\
&k_{ij}=\left|\frac{\partial f_{ij}}{\partial r_{ij}}\right|=\left|\frac{4\varepsilon}{r_{ij}} \left[ \left(156 \frac{\sigma}{r_{ij}}\right)^{12}-42\left(\frac{\sigma}{r_{ij}}\right)^{6}\right]\right|
\end{aligned}
\end{equation}

\begin{figure} [htb]
\begin{center}
		\includegraphics[width=0.70\textwidth]{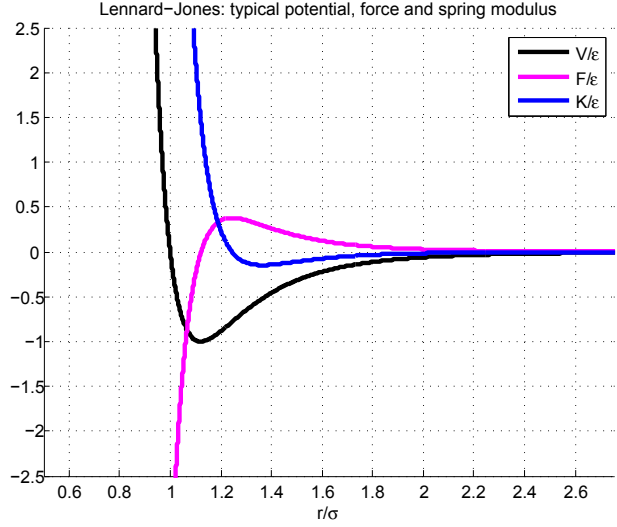}
	\caption{Aspect of the potential, force and spring moduli for a typical Lennard-Jones potential.}
	\label{fig:pot-forc-k}
\end{center}
\end{figure}

The aspect of the results for a typical Lennard-Jones are shown in Fig.
\ref{fig:pot-forc-k}. The force and string modulus for the TLS are derived
directly in each region from the corresponding potential using Eq.
\eqref{eq:f-lj-2} and Eq. \eqref{eq:k-lj}, respectively.

If we accept the expression \eqref{eq:spr-lj} as representative for the
interaction between every pair of atoms of the atomic system, we can consider
the following expression:
\begin{equation}
\label{eq:kuf}
\begin{aligned}
\textbf{K}\textbf{U}=\textbf{F} 
\end{aligned}
\end{equation} 
 
\noindent where $\textbf{U}$ is the displacement vector for all the atoms of
the system. That is: it is a vector of size $n_{at}\ast n_{dm}$, where $n_{at}$ is the
number of atoms and $n_{dm}$ the dimension of the system. Hence, if we consider
the atom at position $i$ and ($i_x$,$i_y$,$i_z$) indicates each space
direction, then the displacement of the atom $i$ is given by
\begin{equation}
\label{eq:ui}
\begin{aligned}
\textbf{U}_{i}=\textbf{u}_i=\begin{vmatrix} u_{i_x}\\u_{i_y}\\u_{i_z}\end{vmatrix}
\end{aligned}
\end{equation} 

The components of the force vector $\textbf{F}$ are built adding the     %
force terms of each possible pair of atoms                                      %

\begin{equation}
\label{eq:Fi}
\begin{aligned}
\textbf{F}^{e}_{i}&=\textbf{f}_{ij}=\begin{vmatrix} f_{ij_x}\\f_{ij_y}\\f_{ij_z}\end{vmatrix}\\
\textbf{F}^{e}_{j}&=\textbf{f}_{ij}=-\begin{vmatrix}f_{ij_x}\\f_{ij_y}\\f_{ij_z}\end{vmatrix}
\end{aligned}
\end{equation} 

Finally the stiffness matrix, of size $(n_{at} \ast n_{dm}) \times (n_{at}
\ast n_{dm})$ is formed by adding the elemental stiffness matrix defined for each
pair of atoms $ij$, definition which practically neglects the effect of a zero
value for a pair of atoms with a distance of  $1.244\sigma$.  That is:

\begin{equation}
\label{eq:Kij}
\begin{aligned}
\begin{pmatrix}K^{e}_{ii_{xx}}&K^{e}_{ii_{xy}}&K^{e}_{ii_{xz}}\\K^{e}_{ii_{yx}}&K^{e}_{i_{yy}}&K^{e}_{ii_{yz}}\\K^{e}_{ii_{zx}}&K^{e}_{ii_{zy}}&K^{e}_{ii_{zz}} \end{pmatrix}&= \begin{pmatrix}k_{ij_{x}}&0&0\\0&k_{ij_{y}}&0\\0&0&k_{ij_{z}}  \end{pmatrix}\\
\begin{pmatrix}K^{e}_{ij_{xx}}&K_{ij_{xy}}&K^{e}_{ij_{xz}}\\K^{e}_{ij_{yx}}&K^{e}_{ij_{yy}}&K^{e}_{ij_{yz}}\\K^{e}_{ij_{zx}}&K^{e}_{ij_{zy}}&K^{e}_{ij_{zz}} \end{pmatrix}&= \begin{pmatrix}-k_{ij_{x}}&0&0\\0&-k_{ij_{y}}&0\\0&0&-k_{ij_{z}} \end{pmatrix} 
\end{aligned}
\end{equation}

\noindent The expression \eqref{eq:kuf}, and its spring like physical meaning allows to introduce boundary conditions as in other classical numerical model, for example, as in the finite
element framework. 

If the system is at equilibrium then the PEL shows a minimum state, where the
resultant forces and displacements are zero. While this is not necessarily a
real physical state, it could be considered as an average description of an
atomic system at low temperatures.  If the PEL is not on a minimum state
then there are resultant forces which produce a displacement of the atoms,
$\textbf{U}$, from the initial position to another, likely more stable one. We
can then update the position using the information on  $\textbf{U}$  and
recompute $\textbf{U}$ using \eqref{eq:kuf} where $\textbf{K}$ and
$\textbf{F}$ corresponds to the updated position. We iterate the process until
the maximum of the absolutes values of $\textbf{U}$ is below a tolerance. 
Then we can accept that the system has arrived to equilibrium and no further
displacement occurs. For our example we fix that tolerance as $0.01r_B$.

We implemented two exit conditions for non-convergence of the iterative process.
The first one is limiting the number of iterations, in our case 200,  and the
second one is using the maximum of the absolute values of $\textbf{U}$, as we
can consider that over a certain displacement, in our case $1.5r_B$, the
configuration of the system has changed beyond control.  In either case we
consider that the calculation did not converge and we reject the results. This
exit conditions achieve a considerable gain in computation velocity
at the cost of not being able to deal with some of the most unstable starting
positions.

\vspace{-6pt}

\section{Lattice and impurities}
\vspace{-2pt}

\begin{figure} [htb]
\begin{center}
	\includegraphics[width=0.8\textwidth]{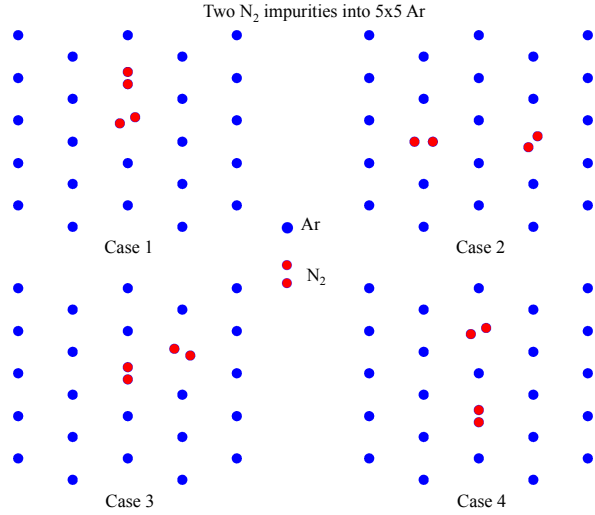}
	\caption{Location of two impurities for checking the lattice finite size effect.}
	\label{fig:2N2}
\end{center}
\end{figure}

We want to minimize the finite size effect on modelization, at least on a
controlled central region, thus if the border influence can be neglected then the
infinite medium assumption for our system can be accepted.  The finite size
effect is directly related to the model size, thus we evaluate it by comparing
different model sizes. We use 5x5, 9x9 and 13x13 pure Ar lattice with hexagonal
structure and the distances corresponding to the experimental crystal
structure~\cite{nielaba}. We analyze the effect of the finite lattice size
considering the presence of impurities. Four different cases with two N$_2$
impurities are modeled. The impurities are placed on different positions but
close to the center region, as shown in Fig.  \ref{fig:2N2} and each N$_2$ is
rotated 30 degrees, independently, to check the possible influence of
orientation. That makes a total of 144 computations for each lattice size.

\begin{figure} [htb]
\begin{center}
	\includegraphics[width=\textwidth]{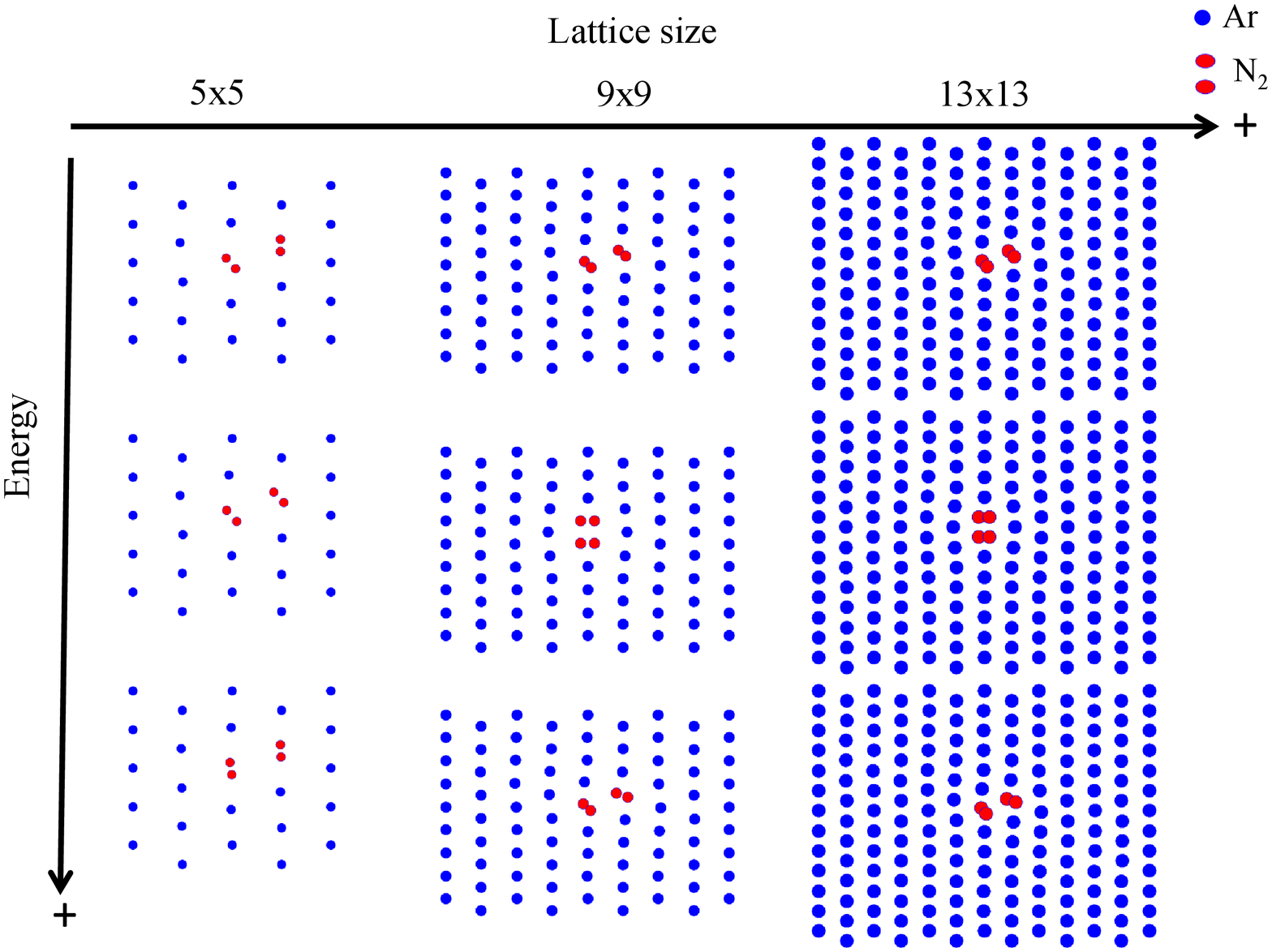}
	\caption{Configurations with the minimum energy for the different lattice sizes.}
	\label{fig:minener}
\end{center}
\end{figure}

The three systems with lowest energy for each lattice size for the former
configurations, which are shown on Fig \ref{fig:minener}, are compared. We can
observe that for the 5x5 lattice the states are not the same as for the 9x9 and
13x13 cases. Moreover the difference of energy between each lowest energy state
for the 9x9 and 13x13 lattice sizes agree. Consequently we can accept that in the
central region of a 9x9 lattice the size effects are negligible. Besides the energy analysis, the
whole computation for the 9x9 lattice took less than 30 minutes on a 2 gigabyte
and  2,53 GHz laptop running under windows 7.  Because of its
feasible computation time, the 9x9 lattice size is a suitable candidate for
statistical studies.

\vspace{-6pt}

\section{ Disordered infinite medium. Excitations for Tunneling effect by TLS-phonon coupling.}
\vspace{-2pt}

The lattice size is chosen as a compromise between avoiding the border effect       
on the central region and having a feasible computationally time.  We start         
from a 9x9 pure Ar lattice with hexagonal structure and the distances               %
corresponding to the experimental crystal structure~\cite{nielaba}. In all our        %
calculations the external Ar frame is kept intact and only the inner 8x8 lattice       %
is relaxed. Similarly, only the inner 8x8 lattice is populated by N$_2$ impurities.

\begin{figure} [htb]
\begin{center}
		\includegraphics[width=\textwidth]{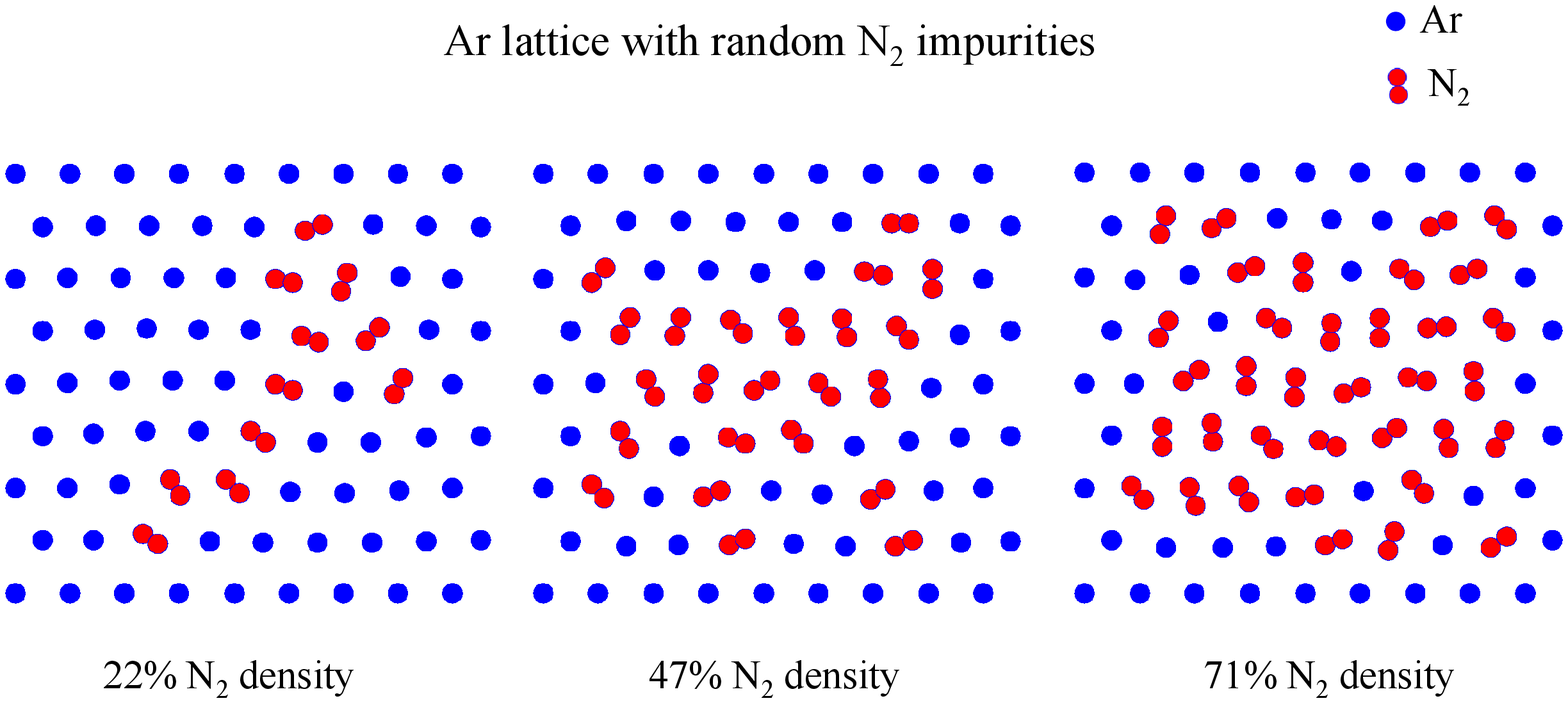}
	\caption{Configurations corresponding to local energy minima obtained as
described in the text for different Ar:N$_2$ ratios.}
	\label{fig:densidades}
\end{center}
\end{figure}

The impurity density is increased iteratively adding one N$_2$ per step.
Particularly,  the concentration of N$_2$ is increased by adding a randomly
located and randomly oriented N$_2$ to the relaxed region of a
previous configuration. For increasing the stability of the system,  the
lattice intersite distance is slightly modified with each N$_2$ addition, in
accordance with the real density of Ar:N$_2$ mixtures. After the addition of
one N$_2$ the system is relaxed.  That procedure simplifies the random
enrichment of the lattices as it starts from a stable configuration and only a
N$_2$ is added in each step. Examples of this process     can be observed for
different densities on Fig. \ref{fig:densidades}.

If the system relaxing converges we save the results, and repeat the procedure
until we have saved 50 stable lattices with the same number of N$_2$
molecules in different positions. For the first N$_2$ added to the
pristine Ar lattice, we  keep the 50 lattices, which we will use as
starting steps for 50 independent iterative procedures. For each
subsequent iteration addition of N$_2$ to each       
of these 50 independent histories, we keep only the lattice with the lowest
energy. This produces 50 independent and relatively low-energy configurations
for each Ar:N$_2$ ratio. That is for each Ar:N$_2$ ratio we check 50x50 stable
configurations taking the lowest energy configuration for each family as
previous configuration for the  following substep.  All the non stable
configurations are discarded. without further computations.  
  
The maximum achieved Ar:N$_2$ ratio is 0.2:0.8, meaning 80\% of inner 7x7 lattice
positions are N$_2$.  To further decrease the energy and approximate the real
ground state, we perform a Monte-Carlo cooling ~\cite{romano,leonel}: for each
state we rotate randomly and sequentially each N$_2$ and keep the new state if
the new configuration converges to a lower energy. We perform 25 sequential
sweeps of each lattice. This decreases the energy an average of $0.14 eV$.

The total number of stable configurations for the 9x9 Ar:N$_2$ were over 97 000
and, including the MC cooling, took about 30 days on a computer with 4 cpus. In
comparison, in~\cite{churkin},  a computer with 200 cpus  need the same time
for a sample of 3000  4x4x4 configurations or 5000 8x8 configurations of
KBr:CN.

We obtained a set of histories for 9x9 Ar lattices with hexagonal structure and
different concentration of N$_2$ randomly placed. The Monte-Carlo cooling
process also  lead the system  to a ground like state. All this process allows
to accept the results as representative of a disordered medium. However, the
goal is the study of the coupling of the different TLSs excitations in a
infinite lattice, therefore only the seven central atoms of the lattice are
considered to minimize border effects. 

\begin{figure} [htb]
\begin{center}
				\includegraphics[width=0.70\textwidth]{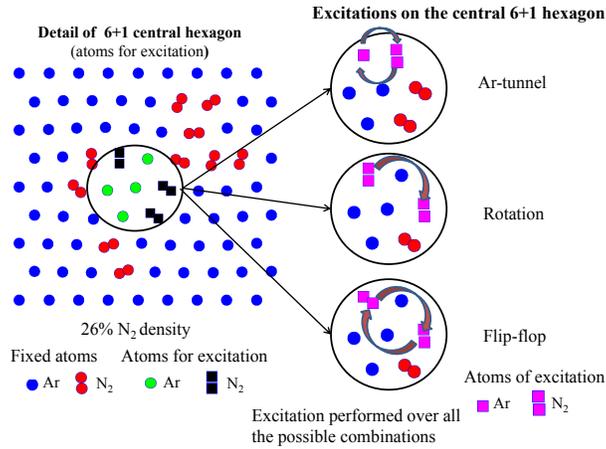}
	\caption{Detail of the excitations}
	\label{fig:excitations}
	\end{center}
\end{figure}

Three types of possible TLSs coupling are considered for this seven central
positions, one which corresponds to a tunneling process where an Ar atom and an
N$_2$ molecule exchange positions.  This we label as Ar-tunnel, the second
excitation corresponds to an orientation change such that an N$_2$ adopts the
orientation of a neighboring, non-parallel N$_2$, and label this as rotation.
The third excitations corresponds to the exchange of orientation of two
non-parallel N$_2$ and is refereed as flip-flop.  An representation of the
central seven atoms and the excitations are shown in Fig.
\ref{fig:excitations}. The excited state is only accepted if (a) its relaxation
process converges and (b) it does not reproduce geometry of its ground state
after the relaxation.

\begin{figure} [htb]
\begin{center}
		\includegraphics[width=\textwidth]{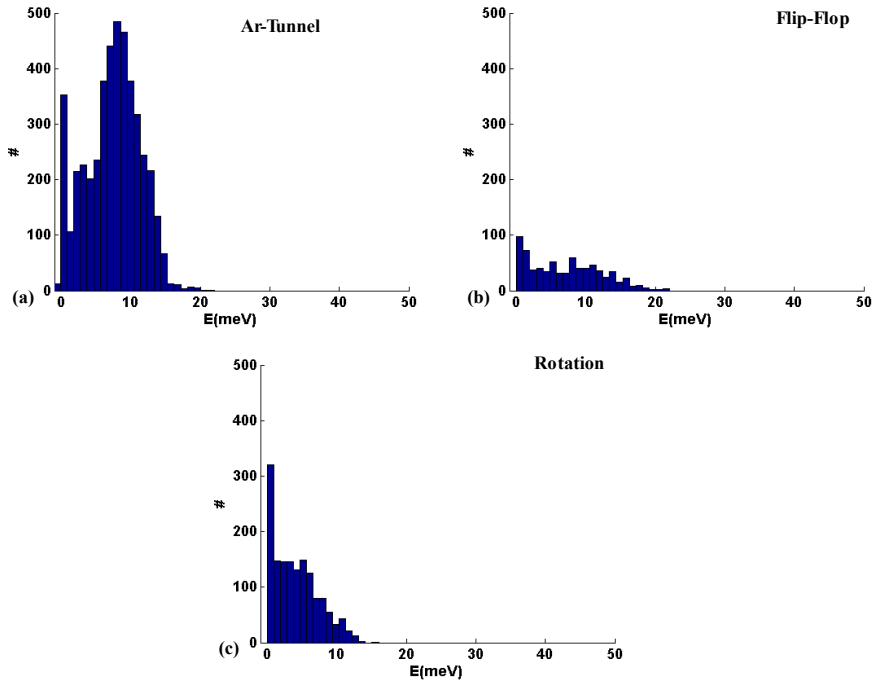}
	\caption{Histograms for TLSs up to 20\% N$_2$ density.}
	\label{fig:TLShist}
\end{center}
\end{figure}

The distribution of TLSs with a N$_2$ density up to 20\% is shown in
Fig. \ref{fig:TLShist}. The comparison of the histograms  allows to identify
which excitations is expected at lower density.  As seen in Fig.~\ref{fig:TLShist}(a), and as expected statistically, Ar-N$_2$ tunneling
process are more likely at low N$_2$ densities compared with processes that
require two N$_2$ molecules. In this case we see a sharp peak of excitations at
near-zero energy, corresponding to the cases where there is an isolated N$_2$
molecule with no neighbours or stress in the vicinity. In these cases, the
tunneling processes result in two effectively equivalent configurations. As
soon as there is a nearby impurity, as is often the case (and always at higher
densities), the excitations form a broad band around 10meV. We
can see that we have over 700 possible TLS for each case on this range and
that is enough to be considered as  a statistically representative sample even for a low density study. The results for higher densities and its physical meaning, which are outside the scope of this paper, are further discussed on~\cite{gaitavi}.


On of the material properties that are of main interest and is the he coupling of the different TLSs with phonons.
For a selected number of small 2D fragments, we employed the same high-quality,
high-cost DFT methods presented in~\cite{alex} to extract the values of the
TLS-phonon interaction energy $\gamma$ of the different TLSs and also to
estimate the compression energy per atom. To test the validity of the numerical
approach we are performing, we construct identical lattices which we treat 
with our Lennard-Jones-based method and with DFT (B3LYP, 6-311G).
We find that the sign of $\gamma_S$ obtained by Lennard-Jones is confirmed by
DFT in all cases examined. While correct in sign and order of magnitude, we do
find that Lennard-Jones overestimates the value of $\gamma_S$ by a factor of
2-4. Lennard-Jones underestimates the value of the compression energy of two
neighboring Ar atoms $C$ by up to an order of magnitude compared with
high-level DFT calculations. On the other hand, a smaller basis set such as
3-21G, which is still much more expensive than the method presented here,
results in values that are comparable to LJ for all problems tested.

\begin{table}[h]
\caption{$\gamma_S$(rotation) in different configurations. Problem A: 3x3
pristine Ar lattice with a central N$_2$.  Problem B: 5x5 pristine Ar lattice
with a central N$_2$.  Problem C: 5x5 pristine Ar lattice with three nearest
neighbours N$_2$ in the central line, oriented parallel to each other and
perpendicular to the line defined by their
centers,~\cite{gaitavi}.} 
\begin{center}
\begin{tabular}{c|c|c|c}
                    &   A    &   B    &   C   \\
\hline
$C$ (LJ,meV)        &   1.0  &  0.15  &  0.5  \\
$C$ (DFT,meV)       &   1.8  &  1.6   &  1.7  \\
$\gamma_S$ (LJ,eV)  &  1.44  &  0.60  &  0.96  \\
$\gamma_S$ (DFT,eV) &  0.60  &  0.320 &  0.22  \\
\end{tabular}
\label{DFT-LJ}
\end{center}
\end{table}

The results in Table~\ref{DFT-LJ} were tested by performing extra calculations,
e.g.  with even larger basis sets such as 6-311+G* or in Problem D, defined as
Problem C but with an extra layer resulting in a 7x7 lattice.  Thus, using
basis set 6-311+G* instead of 6-311G in problem A results in a $\gamma_S=0.56$
eV (rather than 0.60), showing that the 6-311G is already a good enough basis
for this problem. The comparison of Problems C and D at LJ and DFT (3-21G)
levels confirms the qualitative results presented above, both on $C$ and
$\gamma_S$.  

\vspace{-6pt}

\section{Conclusions}
\vspace{-2pt}

We have presented a technique for producing data for studying the coupling between local (TLS) and
extended (phonon) interactions in disordered solids that is capable of dealing
with a statistically representative number of large systems i.e. with
negligible border effects. We rely on a discrete atomic model based on the
well known Lennard Jones Potential, to compute and simulate the equilibrium of
an atomic lattice, with randomly introduced impurities. A MC cooling is also
performed to obtain more stable configurations. The results are extracted from
the central region of a large lattice and thus effectively correspond to a
disordered infinite medium. We applied the developed technique for an analysis
of the coupling on a Ar:N$_2$ mixture, as it is a physically intriguing case of
disordered lattice where ample experimental information is available. 
A discussion about the physical properties of this system which employs the presented method is performed in~\cite{gaitavi}
 
The non linear spring-like models have limitations in terms of accuracy and
convergence in comparison with other minimization methods for discrete systems,
as the conjugate gradient used in~\cite{churkin}. However, the notation and
structure of its mathematical matrix expression is very similar to a finite
element  formulation, thus it simplifies the description of boundary conditions
and enhances its versatility. Moreover, as its stability is very sensitive to
initial conditions, in comparison with other more stable optimization methods,
which allows the fast discarding of configurations which would either fail to
converge or take a long time to converge, effectively reducing the computation
time for producing a statistical representative dataset for TLSs analysis.

The technique, even lacking the accuracy of ab initio models or other non
linear techniques as the conjugate gradient, is able to reproduce  the
qualitative behavior of TLSs. Furthermore,  it solves some problems that do not
allow the application of such more accurate methods to the study infinite
disordered solids considering large samples. Besides, the methodology
introduced is also suitable for its applications with  other potentials,  like
Morse, Harmonic or coulomb electric potentials and  even MC techniques for
considering high temperature systems. However as the basic model is very close
to the MD approach, the techniques developed for the interpretation of material
properties on MD framework can also be applied in our model. Another advantage
of the proposed technique is that the obtained atomic equilibrium positions can
be used as starting position for other more accurate techniques. Thus the
statistically representative sample obtained with our approach improves the
usability range and general applicability of the conventional, more expensive
techniques. 

The strategy employed on the random enrichment of the lattice with impurities
shows further advantages. The random introduction of impurities allows the
consideration of amorphous medium but the iteratively approach helps to achieve
higher impurity density in contrast with a pure random enrichment. Moreover,
having several histories with different level of impurity density increases the
data for the statistical analysis. besides as only the possibility with minimum
energy is stored it helps to have always a ground like state, simplifying the
cooling. The technique is also suitable for performing 3D TLS studies in a
feasible time, however some mathematical and physical considerations have to be
taken into account to enforce stability of those systems as they are more
sensitive to initial conditions.  All in all, we present a computational cheap
technique to produce data suitable for statistical analysis for TLSs properties
with advantages relying on its velocity and versatility. 

The procedure presented also is suitable to study the macroscopic properties of an
amorphous solid starting from its microscopic behavior: the discrete model
description relies on atom positions, forces and displacements, and can be
easily extended and simplifies the  introduction of boundary conditions. Of
course, whenever adequate the parameters of the procedure such as lattice size,
distance of equilibrium, number of steps, number of considered families for the
enrichment and MC steps, can be modified to achieve an improved accuracy.

\end{document}